\documentclass[a4paper,11pt]{article}
\pdfoutput=1 

\usepackage{jcappub} 

\usepackage[T1]{fontenc} 

\newcommand{\ltsima}{$\; \buildrel < \over \sim \;$}
\newcommand{\ltsim}{\lower.5ex\hbox{\ltsima}}

\newcommand{\beq}{\begin{equation}}
\newcommand{\eeq}{\end{equation}}
\newcommand{\beqa}{\begin{eqnarray}}
\newcommand{\eeqa}{\end{eqnarray}}

\def\pa{{\partial}}

\def\ha{\frac{1}{2}}

\title{Testing LTB Void Models Without the Cosmic Microwave Background or Large Scale Structure: \\ {\it New Constraints from Galaxy Ages}}

\author[a,b]{Roland de Putter}
\author[c,a]{Licia Verde}
\author[c,a]{Raul Jimenez}

\affiliation[a]{ICC, University of Barcelona (IEEC-UB), Marti i Franques 1, Barcelona 08028, Spain}
\affiliation[b]{IFIC, Universidad de Valencia-CSIC, 46071, Valencia, Spain}
\affiliation[c]{ICREA, Instituci\'{o} Catalana de Recerca i Estudis Avan\c{c}ats}

\emailAdd{rdeputter@icc.ub.edu}

\abstract{We present new observational constraints on inhomogeneous models based on observables independent of the CMB and
large-scale structure. Using Bayesian evidence we find very strong evidence for the homogeneous LCDM model,
thus disfavouring inhomogeneous models. Our new constraints are based on quantities independent of the growth of
perturbations and rely on cosmic clocks based on atomic physics and on the local density of matter.}

\begin{document}
\maketitle
\flushbottom

\section{Introduction}

The discovery of accelerated expansion about fourteen years ago, based on the dimming of distant supernovae (SN) (\cite{Riess98, Perl99}),
has led to a standard model of cosmology ($\Lambda$CDM) in which about $72 \%$ of the energy of the universe is in the form of a cosmological constant (CC)
(or dynamical dark energy). An important assumption in this model is that our universe is homogeneous.

As an alternative to a cosmological constant, which has significant theoretical problems due to the required fine-tuning,
the supernova data could also be explained without a CC if we live close to the center of a large ($\gtrsim 1$ Gpc),
spherically symmetric
void (see \cite{tomita00, goodwinetal99, celerier00} for some of the first such proposals,
while \cite{mofftat95, mustaphaetal97} discussed this idea even before the measurement of accelerated expansion).
The reason is that we cannot distinguish the effect of spatial variations in the geometry from temporal ones in the radial direction
from measurements along the past light-cone alone.
Specifically, in a large void, the local expansion rate gets larger closer to the center of the underdensity, thus causing the
same additional dimming at large distances that would be caused by accelerated expansion in a homogeneous cosmology.

Such void models arguably have larger philosophical and aesthetic problems than the CC.
First, the existence of a void of the required size is extremely unlikely in the standard inflationary scenario
and its generation would probably involve rather unusual early universe physics.
Moreover, there is a large amount of fine tuning associated with the requirement from observation that we find ourselves within
$\sim 1 \%$ of the void's center (\cite{alnesamarz06, alnesamarz07, BNV10, blommort10, kodamaetal10, foremanetal10}),
a grave violation of the Copernican principle. Additional fine-tuning comes from the fact that
the void has to be close to spherical to be consistent with the isotropy that we observe.
Finally, the void as an explanation for cosmic acceleration does not actually solve the cosmological constant
problem. It merely assumes that the CC is equal to zero without providing any explanation (this criticism of course also applies
to most if not all models of dynamical dark energy).

Still, the final say can only come from observation so it is worth comparing
predictions of the void model to recent cosmological data. This is not just interesting
for the sake of understanding the viability of these specific models, but also serves to test
a not often directly tested
pillar of our cosmological framework, the assumption of homogeneity (see e.g.~\cite{clarkmaart10}).

The void is commonly modeled as a Lema{\^i}tre-Tolman-Bondi universe (LTB, \cite{Lemaitre97, Tolman34, Bondi47}),
which describes the most general spherically symmetric universe filled with pressureless matter (dust).
Such models have been extensively used to reproduce the observed supernova magnitudes
and can in fact explain {\it any} Hubble diagram perfectly, provided enough freedom in the void profile
is allowed
(see e.g.~\cite{mustaphaetal97, yooetal08}).
A more trying test of the LTB cosmology arises when additional data sets are added. Many
data sets have by now been considered, including, but not limited to, the primary cosmic microwave background temperature power spectrum
(e.g.~\cite{alnesetal06, ZMS08, Alexetal09, nadatsar11}),
spectral distortions of the cosmic microwave background (CMB) (\cite{caldsteb08}), the kinetic Sunyaev-Zel'dovich (kSZ) effect (\cite{GBH08b, yooetal10, zibinmoss11, bull12, zhangstebbins11}),
the primordial Lithium abundance \cite{regisclarkson12},
the Baryon Acoustic Oscillation (BAO) scale (e.g.~\cite{GBH08, ZMS08}),
and combinations of the above (\cite{MZS11, ZMS08, Alexetal09, GBH09, BNV10, ZGBRL12}).
In particular, both the combination of the CMB with a low redshift Hubble parameter measurement (and with other data sets) (\cite{MZS11}),
and the observational upper limit on the strength of the kinetic SZ effect in the CMB (\cite{GBH08b, yooetal10, zibinmoss11, bull12, zhangstebbins11}), appear
to rule out at least the simplest versions of the void cosmology.

However, many of the previous studies rely either on the CMB or, through the BAO, on large scale structure (LSS),
while it is not universally accepted that these observables are understood well enough in an inhomogeneous universe. Particularly,
solving the evolution of density perturbations in LTB is notoriously difficult,
although significant progress has by now been made (\cite{clarksonetal09, Zibin08, alonsoetal10, dunsbyetal10, nishietal12, februaryetal12}).
Moreover, constraints derived from the CMB power spectrum may depend on implicit assumptions regarding the distribution of radiation
(\cite{regisclarkson12, clarkreg11}).

It is thus worthwhile considering constraints that do not depend on either CMB or LSS, nor on the details of perturbation
evolution in general. For this reason, we will constrain the geometry/expansion of the universe with data complementary to
supernovae, by using ``red envelope'' galaxy ages (\cite{sternetal10, sternetal10b}) as lower bounds on the age of the universe at various redshifts
in the range $z = 0 - 1.9$.
These estimates of oldest galaxy ages in samples of passively evolving, massive, red galaxies have been used previously
to measure the Hubble parameter $H(z)$ as a function of redshift (see \cite{jimloeb02, simonetal05, sternetal10, sternetal10b}).
In fact, this measurement of $H(z)$ has even been used to constrain LTB models in \cite{FLSC10, wangzhang12}.
However, the Hubble determination crucially depends on the assumption that the average formation time of the oldest galaxies in each sample
is homogeneous. Since this is not guaranteed in a void model (see Appendix \ref{app:A}),
an issue which was at least discussed in \cite{wangzhang12},
we cannot safely estimate $H(z)$ from differential age measurements, and use the ages themselves instead.
As it turns out, the supernova magnitudes, combined with a local determination of the Hubble rate, prefer a void in which the age of the universe as a function
of redshift is low compared to these galaxy ages (see also, e.g., \cite{ZGBRL12}) so that the addition of the age estimates places strong constraints on the model.

We thus carry out a joint analysis of these three data sets, considering a void that is asymptotically flat and with homogeneous big bang time.
We derive parameter constraints and,
performing Bayesian model selection,
find that $\Lambda$CDM is significantly favored over LTB models. The data prefer a very low value of the relative matter density in the center
of the void ($\Omega_{\rm in} \leq 0.1$),
and we show that adding a direct measurement of $\Omega_m$ from, e.g., clusters causes additional tension between the data sets,
helping to further disfavor the LTB model. We thus find results consistent with the studies based on CMB and/or LSS, but using an independent method.

We will briefly review the LTB model and introduce our parametrization in section \ref{sec:model}, and describe the four types of data used
in section \ref{sec:data}. We then derive parameter constraints, and compare the model to $\Lambda$CDM
using both the $\chi^2$ statistic and Bayesian evidence comparison, in section \ref{sec:results}. Finally, we summarize and discuss our results in
section \ref{sec:dis}. The dangers of using red envelope ages to estimate the Hubble parameter in void models are discussed
in Appendix \ref{app:A}.
Throughout this article, we will use notation similar to that
in \cite{ZGBRL12} (and previous papers by these authors).










\section{The Model}
\label{sec:model}

The most general spherically symmetric space-time for a universe filled with pressureless matter ($p=0$)
is given by the LTB metric,
\beq
\label{eq:metric}
ds^2 = dt^2 - \frac{A'^2(r, t)}{1 - k(r)} \, dr^2 - A^2(r, t) \, d \Omega^2,
\eeq
where $d\Omega^2 \equiv d \theta^2 + \sin^2 \theta d\phi^2$ is the squared line element in the transverse direction.
Here, $k(r)$ is the (radial) position dependent curvature function and $A(r,t)/r$ a generalized, position dependent scale factor.
Throughout this work, a $'$ denotes a partial derivative with respect to radial coordinate $r$ and a $\dot{\,}$ will
denote a partial time derivative. The LTB metric allows for different rates of expansion in the longitudinal and transverse directions,
quantified by $H_L = \dot{A}'/A'$ (the longitudinal Hubble rate) and $H_T = \dot{A}/A$ (the transverse Hubble rate),
which are not necessarily equal.

The Einstein Equations imply a generalized Friedman equation (see e.g.~\cite{EnqMatt07, Enq08, GBH08}),
\beq
\label{eq:friedman}
H_T^2 = H_0^2(r) \, \left( \Omega_m(r) \left(\frac{A_0(r)}{A(r,t)}\right)^3 + (1 - \Omega_m(r)) \left(\frac{A_0(r)}{A(r,t)}\right)^2 \right),
\eeq
where $H_0(r) \equiv H_T(r, t_0)$ and $A_0(r) \equiv A(r, t_0)$, and $t_0$
is the cosmic time at present.
Once the gauge freedom in
$A(r,t)$ related to rescalings of $r$ is fixed, the model is completely
specified by the two free functions $H_0(r)$ and $\Omega_m(r)$ which are related to the physical matter density $\rho_m(r)$
and the curvature by
\beq
\frac{3}{8 \pi G} \, H_0^2 \Omega_m(r) = \frac{\int dA A^2 \, \rho_m(r)}{\frac{1}{3} A_0^3},
\eeq
i.e.~the average matter density within $r$, and by
\beq
k(r) = H_0^2(r) (\Omega_m(r) - 1) A_0^2(r). 
\eeq

For a given value of $r$, one recognizes Eq.~(\ref{eq:friedman}) as the Friedman equation for a curved, dust-filled universe
with Hubble parameter $H_0(r)$ and matter density relative to critical density $\Omega_m(r)$. For $\Omega_m(r) < 1$, the solution is
\beqa
\label{eq:sol}
A(r,u) &=& \frac{\Omega_m(r)}{2 (1 - \Omega_m(r))} \, \left( \cosh(u) - 1 \right) \, A_0(r) \nonumber \\
t(r, u) - t_{BB}(r) &=& H_0^{-1}(r) \, \frac{\Omega_m(r)}{2 (1 - \Omega_m(r))^{3/2}} \, \left( \sinh(u) - u \right),
\eeqa
where $t_{\rm BB}(r)$ is in general a free function describing the location dependent Big Bang time.
However, we will require the big bang to be homogeneous, setting $t_{\rm BB} (r)$ to a constant, which without loss
of generality we choose to be zero,
\beq
t_{\rm BB} (r) \equiv 0.
\eeq
This condition ensures that the void can be treated as a perturbation of a homogeneous universe, in the
sense that the void vanishes at early times (\cite{Zibin08, Silk77}).

The requirement of a homogeneous big bang restricts the freedom in model space from two free functions $H_0(r)$ and $\Omega_m(r)$
to one free function and one parameter, which we choose to be $\Omega_m(r)$ and $H_{00} \equiv H_0(r=0)$.
Once $\Omega_m(r)$ and $H_{00}$ have been specified,
$t_0$ can be calculated at $r=0$ from Eq.~(\ref{eq:sol}) by first finding $u_0(r=0)$ such that $A(0,u_0(r=0))=A_0(0)$
and then inserting $u_0(r=0)$ into the expression for $t(r,u)$. At any other $r$, requiring $t(r,u_0(r))=t_0$
then determines $H_0(r)$.

In order to calculate observables in the following sections, we will calculate quantities along the past lightcone.
The relation between $t$ and $r$ along a radial null geodesic is obtained by setting the line element (\ref{eq:metric})
to zero,
\beq
\frac{dt}{dr} = \pm \frac{A'(r,t)}{\sqrt{1 - k(r)}}.
\eeq
The photon gravitational redshift in an LTB universe is given by (\cite{Bondi47, EnqMatt07})
\beq
\frac{d\ln(1 + z)}{dt} = - H_L(r,t).
\eeq
We will calculate $r, t$ and any observables of interest as a function of redshift
by integrating
\beqa
\frac{dt}{d\ln(1+z)} &=&  - H_L^{-1}(r,t) \nonumber \\
\frac{dr}{d\ln(1+z)} &=&  \pm \frac{\sqrt{1 - k(r)}}{\dot{A}'(r,t)},
\eeqa
and using Eq.~(\ref{eq:sol}) to find $A(r,t)$ and its derivatives.

Fixing $A_0(r) = r$, we parametrize the void profile $\Omega_m(r)$ by an asymptotically flat, constrained (in the sense that the void is not compensated by an overdensity
at large radius) GBH profile (\cite{GBH08}),
\beq
\label{eq:profile}
\Omega_m(r) = 1 - (1 - \Omega_{\rm in}) \, \left( \frac{1 - \tanh \left[ (r - R)/2\Delta R \right]}{1 + \tanh \left[ R/2\Delta R \right]} \right).
\eeq
The void is then completely described by the parameters $\Omega_{\rm in}$ (the relative matter density in the center of the void), $R$
(the void radius), $\Delta R$ (the width of the transition from void to background) and $H_{00}$ (the expansion rate at the void center).

We will assume our galaxy is at the center of this void. If it were not, the universe would be anisotropic relative to our position.
In particular, the observed cosmic microwave background dipole moment restricts us to be within about $1 \%$ of the void radius from
the center (\cite{alnesamarz06, alnesamarz07, BNV10, blommort10, kodamaetal10, foremanetal10}, but see \cite{regisclarkson12, clarkreg11} for a potential caveat
related to the non-negligible effects of radiation).

We have ignored the contribution from radiation in the above. Since we only consider low redshift observables, at times where the
radiation component would be negligible, we expect this to be a good approximation.

We will compare the void model described above to a homogeneous, spatially flat $\Lambda$CDM universe, filled with pressureless matter and a cosmological constant.
Thus, the number of $\Lambda$CDM parameters relevant for the observables considered in this work is two ($\Omega_m$ and $H_0$), while in the void model, we vary four parameters.

\section{Data}
\label{sec:data}

It is well known that void models can be fine-tuned to provide a good fit (in comparison with $\Lambda$CDM) to supernova luminosities.
In addition, there is enough remaining freedom to provide a good fit to a low redshift determination of the Hubble parameter.
We will therefore use the Union 2.1 compilation of supernova data \cite{union21} and the measurement of the local expansion rate by \cite{hst11} as our
base set of data to compare the void model against. Tension between the void model and the data arises when more data sets are added.
The main new ingredient of this work is the realization that lower bounds on the age of the universe at different redshifts, as obtained from
red envelope galaxies (see \cite{jimloeb02, simonetal05, sternetal10, sternetal10b} and references therein), have strong constraining power. In addition,
we will consider the effect of a measurement of the local matter density.
We specifically avoid using measurements of quantities that depend on the details of cosmic perturbations.
We describe the four data sets/compilations and their likelihood functions
in more detail below. In the next section, we will compare these data against the void model and $\Lambda$CDM predictions,
using the independence of the different data sets to construct the total likelihood
as the product of the individual likelihoods.

\subsection{Type IA supernovae}
\label{subsec:sn}

We use the recent Union 2.1 compilation of Type IA supernovae (\cite{union21}),
which contains 557 supernova magnitudes in the redshift range $z = 0.015 - 1.4$.
The expectation value of the observed (stretch-calibrated) magnitude at the peak of the lightcurve can be written
(we use hats to identify observables/estimators)
\beqa
\label{eq:sn mag}
\langle \hat{m}_i \rangle &=& M - 5 \log_{10}\left(c^{-1} H(0, t_0) [{\rm Mpc}^{-1}]\right) + 25
+ 5 \log_{10}\left(c^{-1} H(0, t_0) d_L(z_i)\right) \nonumber \\
&\equiv& \mathcal{M} + \mu_i(\theta),
\eeqa
where we have defined the distance modulus
\beq
\mu_i(\theta) \equiv 5 \log_{10}\left(c^{-1} H(0, t_0) d_L(z_i)\right),
\eeq
which depends on cosmological parameters (and model) $\theta$ (we have suppressed the parameter dependence in some places to avoid clutter).
Above, $M$ is the absolute supernova magnitude, which, after calibration, is assumed to be universal,
\beq
d_L(z_i) = (1 + z)^2 \, A(r(z), t(z))
\eeq
is the luminosity distance to the redshift $z_i$ of the $i$-th supernova
and $H(0, t_0)$ is the Hubble rate at $z = 0$ and $r = 0$, and is equal to $H_0$ ($H_{00}$) in the homogeneous (inhomogeneous) case.

Assuming the observed magnitudes $\hat{m}_i$ are drawn from a Gaussian distribution with means given by
Eq.~(\ref{eq:sn mag}) and with covariance matrix ${\bf C}_{i j}$ (we use the matrix {\it with} systematic errors provided in the Union 2.1 compilation),
marginalization over the nuisance parameter $\mathcal{M}$ (see for instance Appendix F of \cite{LewisBridle02}) leads to
\beq
\chi^2_{\rm SN} = \sum_{ij} {\bf C}^{-1}_{i j} (\hat{m}_i - \mu_i(\theta))  (\hat{m}_j - \mu_j(\theta)) - \frac{\left[ \sum_{ij} {\bf C}^{-1}_{i j} (\hat{m}_i - \mu_i(\theta)) \right]^2}{\sum_{ij} {\bf C}^{-1}_{i j}}.
\eeq
The normalization is such that $\langle \chi^2 \rangle = N_{\rm SN}$.
Because of overlap between the Union sample with the data set used for the Hubble parameter described in the next subsection, we omit
all supernovae at $z < 0.1$ from the Union sample to avoid double-counting of information. After this cut, $N_{\rm SN} = 392$ supernovae remain.

\subsection{The local expansion rate}
\label{eq:h0}

We use the measurement of the Hubble parameter $H_0$ by \cite{hst11} who used Hubble Space Telescope (HST) data. This work makes use of Cepheids to construct a distance ladder to
eight nearby Type IA supernovae, allowing a determination of the absolute supernova magnitude. Combining this with information from a Hubble
diagram of supernovae at $z < 0.1$, $H_0$ can be calculated, leading to $\hat{H}_{0} = 73.8$ km $s^{-1}$ Mpc$^{-1}$
with an error
$\sigma_{H_{0}} = 2.4$ km $s^{-1}$ Mpc$^{-1}$.

This determination of the expansion rate assumes a cosmology to calculate the distance moduli in the supernova
Hubble diagram. In $\Lambda$CDM, the potential error in $H_{0}$ due to assuming an incorrect cosmology is small because at low redshift the distance moduli
do not have a strong cosmology dependence.
However, since we study cosmology beyond $\Lambda$CDM, we consider it more prudent to treat the HST measurement as a measurement of the (inverse)
luminosity distance to the effective redshift of the
supernova sample used for the Hubble diagram (see also \cite{Percetal10}). The measurement of $H_0$ given above corresponds to a measurement of
$\hat{d}^{-1}_L(z_{\rm eff}) = 5.97 \cdot 10^{-3}$ Mpc$^{-1}$
with
$\sigma_{d^{-1}_L} = 1.9 \cdot 10^{-4}$ Mpc$^{-1}$,
where $z_{\rm eff} = 0.04$.
The $\chi^2$ is
\beq
\chi^2_{H_0} = \left(\frac{\hat{d}^{-1}_L - d^{-1}_L(\theta)}{\sigma_{d^{-1}_L}} \right)^2.
\eeq


\subsection{Galaxy ages}
\label{subsec:ages}

Age measurements of massive, red galaxies
can be used to estimate the upper edge of the age distribution at each redshift, the so-called red envelope ages.
These measurements of the oldest galaxy ages vs. redshift can be used as a redshift-dependent lower bound on the age of the universe.
In total we use 32 such age estimates in the redshift range $z = 0.1 - 1.85$, with independent error bars at the $10 \%$ level (see Fig.~\ref{fig:ages}).
We refer to \cite{simonetal05, sternetal10, sternetal10b} and references therein for details on the data sets used and on the age estimation
from galaxy spectra.

The likelihood of a set of (envelope) age measurements $\hat{a}_i$ (with uncertainty $\sigma_i$), given a cosmological model $\theta$, is given by
\beq
p(\{\hat{a}_i \} | \theta, \{t_{f,i} \}) \propto {\rm exp}\left[ - \ha \sum_i \sigma_i^{-2} \left( t_i(\theta) - (\hat{a}_i + t_{f,i}) \right)^2 \right],
\eeq
where we have not written the normalization factor, as it is independent of the cosmology and therefore will not be relevant
for parameter constraints and model comparison.
The label $i$ runs over all 32 redshift in our data set, $t_i(\theta)$ is the cosmic time
at redshift $z_i$ in the cosmology $\theta$,
and $t_{f,i}$ is the formation time of the red envelope galaxy at each position.
In a homogeneous universe, one can assume this formation time to be independent of position, $t_{f,i} = t_f$, so that differences in ages between different redshifts
can be used to estimate $H(z)$ as in \cite{simonetal05, sternetal10, sternetal10b}. However, in an inhomogeneous cosmology,
the formation time may depend on position within the void, which could significantly bias the resulting Hubble measurement. We discuss this issue in more
detail in Appendix \ref{app:A} (see also \cite{wangzhang12}).
As a consequence, we need to treat the formation time at each position as an independent, free parameter\footnote{Alternatively, if we could model the formation time's position dependence $t_f(r)$,
the lost information on the Hubble rate can be restored, and information may even be added through potential cosmology dependence of $t_f(r)$.
We will not pursue this approach here, as it will require modeling of structure formation, which we wish to avoid in this work.}.

Demanding only that the formation times $t_{f,i} \geq 0$, and assuming a uniform prior,
allows us to analytically marginalize over the formation times,
\beqa
p(\{\hat{a}_i \} | \theta) \, &\propto& \, \prod_i \int_0^\infty d t_{f,i} \, \,  p(\hat{a}_i | \theta, \{t_{f,i}\}) \nonumber \\
&\propto& \, \prod_i \left(1 - {\rm erf}\left( \frac{\hat{a}_i - t_i(\theta)}{\sigma_i} \right)\right).
\eeqa
This then gives us an effective $\chi^2$ of
\beq
\chi^2_{\rm Ages} = -2 \ln{p(\{\hat{a}_i \} | \theta)} = -2 \sum_i \ln\left[1 - {\rm erf}\left( \frac{\hat{a}_i - t_i(\theta)}{\sigma_i} \right) \right],
\eeq
where we again omit cosmology independent (additive) terms.



\subsection{The local matter density}
\label{subsec:locmat}

We use a measurement of the local matter density at redshift zero by \cite{bahcalletal00},
which we can treat, in the context of LTB models, as a measurement of $\Omega_m(0)$ (equal to the parameter $\Omega_{\rm in}$
in the GBH parametrization and simply $\Omega_m$ in $\Lambda$CDM).
Using the mass-to-light ratio of galaxy clusters, they find
$\hat{\Omega}_m = 0.16$ with uncertainty $\sigma_{\Omega_m} = 0.06$. We thus use the following $\chi^2$,
\beq
\chi^2_{\Omega_m} = \left( \frac{\hat{\Omega}_m - (\Omega_m(0))(\theta) }{\sigma_{\Omega_m}} \right)^2.
\eeq

More recent measurements of $\Omega_m$ exist that give significantly smaller error bars (and values close to
$\Omega_m = 0.28$). However, these more recent determinations typically depend on the assumption of a homogeneous cosmology
and often rely on a good understanding of the cosmic microwave background or the distribution of large scale structure.
We choose the measurement by \cite{bahcalletal00} because it depends largely on well understood local physics. It was derived by using the
observed
mass-to-light ratios of galaxy clusters to estimate the mass-to-light ratio of the local universe.
This quantity was then multiplied by the observed total luminosity of the local universe to obtain the
local matter density (we expect the cosmology dependence of the conversion factor between observed cluster mass-to-light ratio
and total mass-to-light ratio, which was derived from simulations, to be small compared to the error bar).

\section{Results}
\label{sec:results}

We now derive constraints from the data sets described in Section \ref{sec:data} (abbreviated as SN, $H_0$, Ages, $\Omega_m$),
paying particular attention to model comparison between the void model
and $\Lambda$CDM. For this purpose,
we consider two statistics. First, we will compare the difference in best-fit $\chi^2$ values,
corrected for the number of degrees of freedom,
\beq
\chi^2_{\rm void} - \chi^2_{\rm \Lambda CDM} - ({\rm dof}_{\rm void} - {\rm dof}_{\rm \Lambda CDM}) \equiv \Delta \chi^2 - \Delta {\rm dof},
\eeq
with $\Delta {\rm dof} = 4 - 2 = 2$.
Since neither model is nested inside the other, it is difficult to turn this statistic into a useful quantitative diagnostic to
choose between the two models.

\begin{figure}[htb!]
\centering
\includegraphics[width=\linewidth]{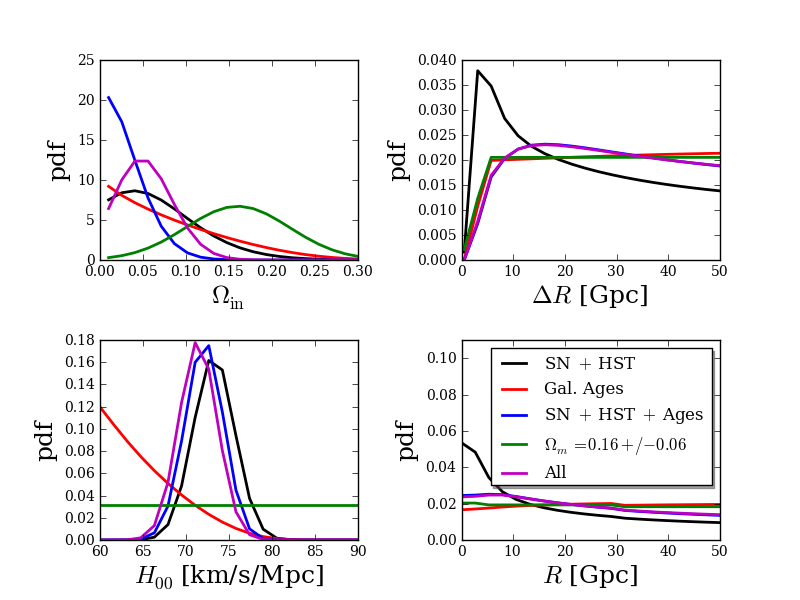} 
 \caption{Marginalized posterior probability distributions of the void model parameters.
  }
  \label{fig:1dpdf}
\end{figure}

Secondly, we will consider the Bayes factor, see e.g.~\cite{jeffreys61, mackay03},
\beq
B \equiv \frac{E_{\rm \Lambda CDM}}{E_{\rm void}},
\eeq
where the Bayesian evidence is defined as
\beq
E \equiv \int d\theta \, p(\theta) \, p(\hat{D} | \theta),
\eeq
and can be seen as the average likelihood of the data over a model's parameter space.
Here, the integral is over the parameter space of the model, $p(\theta)$ is a prior on the model parameters $\theta$,
and $p(\hat{D} | \theta)$ the likelihood given the data $\hat{D}$.
The Bayes factor can be interpreted in terms of betting odds when the models are deemed a priori equally likely, i.e.~the odds are $B:1$ in favor of $\Lambda$CDM,
and we will interpret it using the scale suggested by \cite{jeffreys61} (see also \cite{kassraftery95}), which we show in Table \ref{tab:jeffreys}.
Note that, while in the majority of cosmological applications the Bayes factor has been used
to compare nested models, this statistic is of straightforward use and
interpretation also for non-nested models, as is the case here.

\begin{table*}[hbt!]
\begin{center}
\small
\begin{tabular}{c|c}
\hline\hline
$B$ & evidence against void model\\
\hline\hline
1 to 3.2 & barely worth mentioning\\
3.2 to 10 & substantial\\
10 to 32 & strong\\
32 to 100 & very strong\\
$>$ 100 & decisive\\
\hline\hline
\end{tabular}
\caption{Jeffreys' scale for interpreting the Bayes factor.
}
\label{tab:jeffreys}
\end{center}
\end{table*}

The Bayes factor strongly depends on the priors $p(\theta)$.
The (relevant part of) $\Lambda$CDM parameter space consists of $\Omega_m$ and $H_0$ and we will put the same uniform priors on these parameters as on
their void model counterparts $\Omega_{\rm in}$ and $H_{00}$. We choose the prior ranges on these parameters such that
all significant likelihood is included within them, so that the ratio of evidences becomes independent of these ranges.
The other void parameters $R$ and $\Delta R$ do not have a $\Lambda$CDM counterpart and we choose priors
$R = 0 - 50$ Gpc and $\Delta R = 0.5 - 50$ Gpc, with an additional requirement that $\Delta R > 0.1 R$ for reasons of numerical stability
(this latter requirement does not have a significant effect on the final probability distributions or Bayes factors).

\begin{figure}[htb!]
\centering
\includegraphics[width=\linewidth]{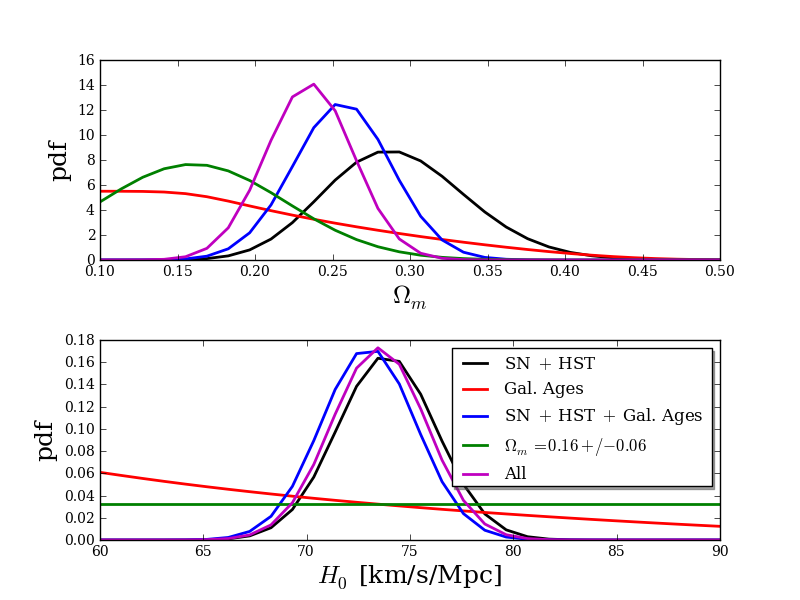} 
 \caption{ Marginalized posterior probability distributions of the $\Lambda$CDM parameters.
  }
  \label{fig:1dpdflcdm}
\end{figure}

\subsection{SN + $H_0$}

Considering first the supernova and local Hubble rate measurements, both models provide a good fit, as
expected. Comparing the $\chi^2$'s at the best-fit points in the two parameter spaces, we find
$\Delta \chi^2 - \Delta {\rm dof} = 1.6$, not strongly favoring either model.
The Bayes factor, on the other hand, equals $B = 13.2$, thus (just) being in the range considered ``strong'' evidence
for $\Lambda$CDM over the void cosmology, according to Jeffreys' scale.
The reason that the Bayesian analysis more strongly disfavors the void model than the $\chi^2$ based analysis
is that the Bayes factor penalizes models that need a large parameter space to obtain a good fit. The void model has two parameters more than
$\Lambda$CDM, both ($R$ and $\Delta R$) being poorly constrained.

The black curves in Fig.~\ref{fig:1dpdf} show the marginalized probability distributions of the parameters
in the void model (see Fig.~\ref{fig:1dpdflcdm} for the $\Lambda$CDM case, also for the curves discussed in the following subsections).

\begin{figure}[htb!]
\centering
\includegraphics[width=\columnwidth]{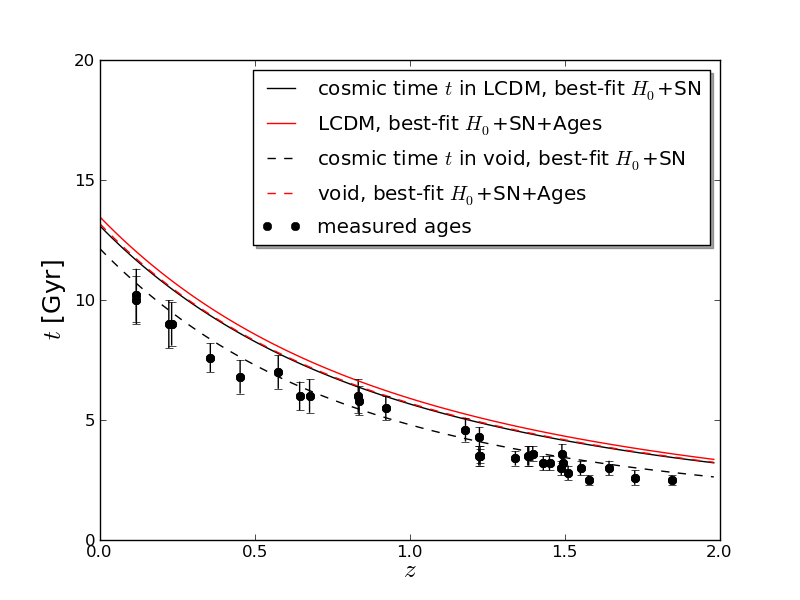} 
 \caption{Age of the universe versus redshift in void model and $\Lambda$CDM, compared to observed lower bounds from galaxy ages.
 When the void model parameters are chosen to provide an optimal fit to the supernova and local Hubble parameter measurements,
 the resulting cosmology is in significant conflict with the age observations (dashed black curve).
 $\Lambda$CDM, on the other hand, is consistent with the age data when the SN $+ H_0$ best-fit
 parameters are used (solid black).
 When the void model parameters are adjusted to agree with the age data and SN $+ H_0$ simultaneously,
 a much better fit to the age data can be reached (red dashed), but both the best-fit $\Delta \chi^2$ statistic relative to $\Lambda$CDM
 and the Bayesian evidence change in favor of $\Lambda$CDM when these three data sets are combined (see text).
  }
  \label{fig:ages}
\end{figure}

\subsection{Adding galaxy ages}

In Fig.~\ref{fig:ages}, we show the age of the universe as a function of redshift
for the SN + $H_0$ best-fit points in parameter space for both the void model and $\Lambda$CDM.
Also plotted are the galaxy based red envelope ages described in Section \ref{subsec:ages},
which provide a lower bound on the age of the universe.
One sees that while the $\Lambda$CDM model (solid black line) is perfectly consistent with these observations,
the void model (dashed black) is under some tension. More quantitatively, the difference in $\chi^2_{\rm Ages}$
between the two models (for these specific parameter values)
is
$\Delta \chi^2_{\rm Ages} = 48$ (specifically, $\chi^2_{\rm Ages} =  49.9 (2.3)$ for LTB ($\Lambda$CDM)). This difference in $\chi^2$ is significantly alleviated
once we allow the model parameters to vary (red curves) and we again compare best-fit $\chi^2$ values.
The same tension can also be seen by comparing the black, SN + $H_0$, curves in Fig.~\ref{fig:1dpdf} to the red ones,
which shows the parameter distribution preferred by the galaxy age data alone. Clearly, the galaxy ages prefer lower values of
$H_{00}$ (and $\Omega_{\rm in}$) than the SN + $H_0$ data, and require larger void dimensions.

To get a complete picture of the effect of the galaxy age data, we next calculate the likelihoods for the full parameter spaces
with the combined data set SN + $H_0$ + Ages. The resulting parameter constraints are shown by the blue curves of Fig.~\ref{fig:1dpdf}.

Both models still provide a good fit to the data in the sense that the best-fit $\chi^2$ values are compatible with the number of observables minus number of degrees of freedom.
However, the {\it difference} between the two models in best-fit $\chi^2$'s is now
$\Delta \chi^2 - \Delta {\rm dof} = 5.4$, showing that the void model is quite strongly disfavored relative to $\Lambda$CDM also based on the $\chi^2$
statistic. 
Comparing the Bayesian evidences gives
$B = 17$, strengthening the evidence against the void model.

\subsection{Adding the local matter density}

As can be seen in Fig.~\ref{fig:1dpdf}, the combination of data discussed in the previous subsection requires a very low value of
$\Omega_{\rm in}$ in the void model. However, measurements of the local matter density exist that point at a larger value, more consistent with
the preferred $\Lambda$CDM value of $\Omega_m$. Thus, using such measurements, the void model can be further disfavored relative to $\Lambda$CDM.
As a proof of concept, we use the measurement described in Section \ref{subsec:locmat}, even though this measurement is rather old and its value
is even low compared to $\Lambda$CDM (see Section \ref{subsec:locmat} for a discussion).

\begin{figure}[htb!]
\centering
\includegraphics[width=\columnwidth]{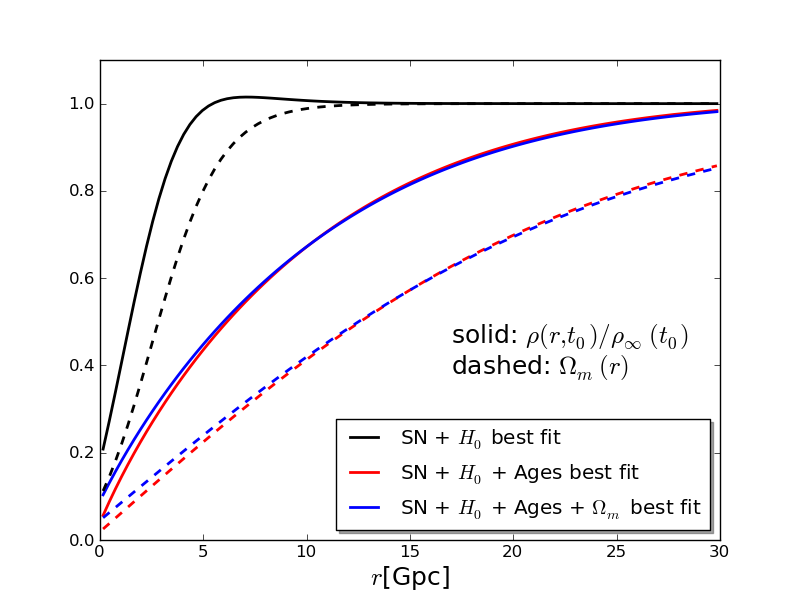} 
 \caption{The best fitting density profiles to the three different data combinations considered in this section.
 The solid lines show the present physical matter density relative to the matter density outside the void as a function of radial position $r$.
 The dashed lines show the quantity $\Omega_m(r)$ (see Eq.~(\ref{eq:profile})), which gives the average matter density {\it within}
 $r$ relative to a critical density defined in terms of the transverse Hubble parameter at $r$.
 While the supernova and $H_0$ data allow for a void of ``only'' a few Gpc in radius, the galaxy age data prefer a larger void,
 several tens of Gpc in radius, although the size distribution is broad (see e.g.~Fig.~\ref{fig:2dpdf}).
  }
  \label{fig:profiles}
\end{figure}

We show the parameter probability distributions preferred by this measurement alone in green in Fig.~\ref{fig:1dpdf} (of course, it only constrains
$\Omega_{\rm in}$ and $\Omega_m$) and the combined constraints from SN + $H_0$ + Ages + Local Matter in magenta.
The value of the best-fit $\Delta \chi^2 - \Delta {\rm dof}$ now equals $8.0$ (although both models are still able to provide a good fit to the data in terms of the absolute
values of the best fit $\chi^2$'s) and the Bayes factor is
$B = 39.8$, making the evidence in favor of a homogeneous universe ``very strong''
according to Jeffreys' scale.

\begin{figure}[htb!]
\centering
\includegraphics[width=\linewidth]{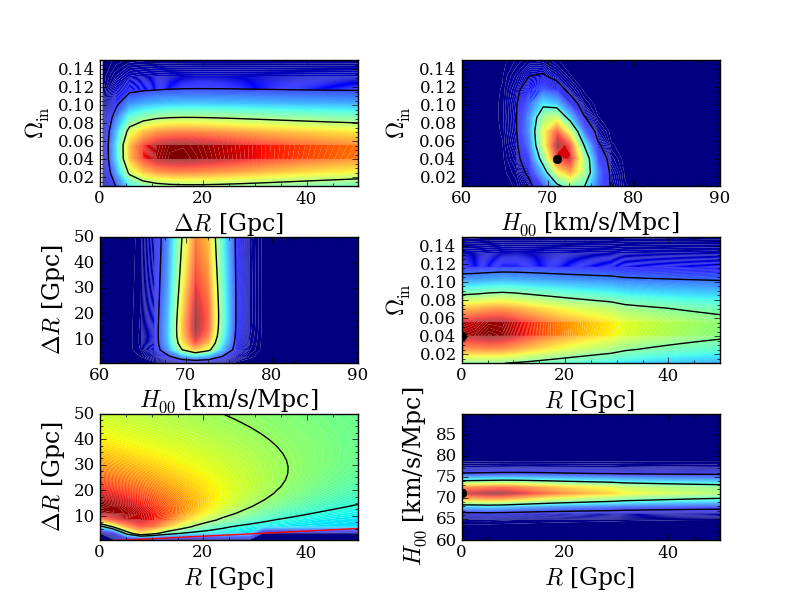} 
 \caption{ Marginalized two dimensional posterior probability distributions of the void model parameters.
 Black contours indicate $68 \%$ and $95 \%$ confidence level regions. The black dot indicates the best-fit point.
 The region below the red line in the lower part of
 the $R - \Delta R$ plane is excluded by our prior.
  }
  \label{fig:2dpdf}
\end{figure}

The best-fit void profiles for the different combinations of data sets are shown in Fig.~\ref{fig:profiles}.
We show the final two dimensional $68 \%$ and $95 \%$ confidence level contours in Fig.~\ref{fig:2dpdf}. As can also be seen from
Fig.~\ref{fig:1dpdf},
the parameters governing the void dimensions,
$R$ and $\Delta R$,
are clearly poorly constrained. Even with our broad priors, with upper limits of $50$ Gpc, we do not enclose all the likelihood within the prior ranges.
However, we do not consider even broader priors for computational reasons. We do note that if we broadened the priors, the evidence for the void model
would decrease, favoring $\Lambda$CDM even more strongly than with the current prior. By tightening the prior ranges in $R$ and $\Delta R$, it is in principle possible
to improve the Bayesian evidence for the void model and thus make the Bayes factor smaller. However, optimizing priors to get a more favorable
result is against the spirit of Bayesian analysis, which is why we do not quote these numbers. We did check that we can at most get a $30 \%$ increase
in the evidence, and thus about a $30 \%$ decrease in the Bayes factor. Therefore, optimizing the prior ranges would not cause a significant change in
our main conclusions.

\section{Discussion and conclusions}
\label{sec:dis}

In this work, we have constrained LTB models proposed to explain the apparent accelerated expansion of the universe
without introducing a cosmological constant. We have tested the LTB model, together with $\Lambda$CDM, against several types of data,
including a recent compilation of supernova data, and the strongest constraint to date on the local expansion rate.
New to this analysis is the addition of age data of old, passively evolving galaxies. We have argued that, while it is not allowed
to use the differential ages as a Hubble parameter measurement in the context of void models, the data still have strong constraining power
when considered as a set of lower bounds on the age of the universe in a wide redshift range $z \approx 0 - 1.9$. Finally, we added
a cluster-based measurement of the local (relative) matter density $\Omega_m$ to further constrain the models.

In the previous section, we found that, while in terms of goodness of fit, both the void model and $\Lambda$CDM provide a satisfactory fit
to all data combinations considered, a comparison of the relative merits of each model strongly favors the latter.
Specifically, $\Lambda$CDM provides a better best-fit $\chi^2$, with the difference
($\Delta \chi^2 = \chi^2_{\rm void} - \chi^2_{\rm \Lambda CDM}$), after correcting for the fact that the void model
has two more free parameters, being
$\Delta \chi^2 + 2 = 1.6, 5.4$ and $8.0$ for SN + $H_0$, SN + $H_0$ + Ages and SN + $H_0$ + Ages + $\Omega_m$ respectively.
Bayesian analysis provides a good way of comparing the two models against the data through the Bayes factor, which is the ratio of
Bayesian evidences. In addition to how well each model is able to fit the data, the advantage of this statistic is that it also takes into account
Occam's Razor by penalizing models that need a large parameter space to obtain a good fit. The Bayes factors (here, the ratio of $\Lambda$CDM over void evidence)
are $B =  13.2, 17.0$ and $39.8$ for the data combinations listed above. Therefore, the first two combinations of data sets provide
``strong'' evidence against the void model according to Jeffreys' scale, while the combination of all data even gives ``very strong'' evidence.

Our results are consistent with studies based on the kinetic SZ effect \cite{GBH08b, yooetal10, zibinmoss11, bull12, zhangstebbins11}, which find
that void models that can fit the supernova data are ruled out because they predict a much larger kSZ signal than consistent with observation (although this does
assume the matter and radiation perturbations are adiabatic), and with studies combining the CMB temperature spectrum with other data sets.
For instance, \cite{MZS11} shows, among other things, that the CMB in combination with supernovae require a very low value of $H_{00}$, clearly in conflict with the
measurement by \cite{hst11}, and that void models predict a very low clustering amplitude at low redshift, which is also inconsistent with data from clusters and other probes.
The novel thing about our work is that it provides independent confirmation of the problems with large void models, without using CMB or LSS data.
Moreover, our study suggests that combining the age data we used, with the data used in other studies, including the CMB and LSS data,
could lead to significantly larger Bayes factors, even more firmly ruling out the void model. Of course, in that approach one would lose the independence of
the details of cosmic perturbations in void models, which was one of the advantages of our approach in this work.

Large values of the parameters governing the void size, $R$ and $\Delta R$, are not well constrained by our data, with void sizes $\gtrsim 50$ Gpc
still allowed. These parameters have much stronger upper limits when CMB and LSS data are included. For instance, using
SN + $H_0$ + BAO + CMB data, \cite{ZGBRL12} finds $R$ and $\Delta R$ to be at most a few Gpc, with $R = 0.18^{+0.64}_{-0.18}$ Gpc and
$\Delta R = 2.56^{0.28}_{-0.24}$ Gpc at $68 \%$ confidence level.

There are some potential caveats to the analysis presented here. First of all, we restricted ourselves
to asymptotically flat cosmologies, $\Omega_{\rm out} = 1$. In addition, we assumed a homogeneous big bang time such that
at early times, the void is just a small perturbation onto a homogeneous Einstein-de Sitter cosmology. Finally,
we did not consider the most general void profiles, instead relying on a specific parametrization involving four free parameters.
Specifically, we did not include void profiles of the ``unconstrained'' type, which (partially) compensate the central underdensity with an overdensity
near the edge of the void.

Our study thus strongly constrains the class of the simplest, perhaps most reasonable LTB models. Relaxing some or all of the above assumptions might improve
how well the model fits the data, but also adds more parameter space. It is thus not clear if this would strengthen or weaken the evidence against
LTB cosmologies. Our main goal here was to introduce an alternative, strong method to constrain void models, and a study of constraints
on the most general void cosmologies is beyond the scope of this article.

In conclusion, void models as an explanation for the apparent cosmic acceleration are under attack from a number of independent types of
cosmological data and are by now looking less and less like a viable alternative for a cosmological constant or dark energy.





\acknowledgments

RdP and LV are supported by FP7-IDEAS-Phys.LSS 240117. LV and RJ are supported by FPA2011-29678-C02-02.

\appendix
\section{Inhomogeneous galaxy formation time}
\label{app:A}

The galaxy age data used in this work (see section \ref{subsec:ages}) have in the past been used to estimate the Hubble parameter as a function
of redshift. This method works because the passively evolving galaxy samples employed to determine the ``red envelope'' ages
can in a homogeneous universe be assumed to have formation times whose statistics do not depend on position (i.e.~on the redshift at which they are observed).
The age difference between different redshifts then provides an unbiased estimator of the difference in cosmic time between those redshift,
allowing a Hubble measurement through the relation
\beq
H^{-1} = - \frac{dt}{d\ln(1+z)}.
\eeq
In an inhomogeneous cosmology, this relation becomes
\beq
H^{-1}_L = - \frac{dt}{d\ln(1+z)},
\eeq
so that a differential cosmic time measurement could still carry valuable information.
The problem with using ages is that, in an inhomogeneous universe, the formation time corresponding to red envelope ages at different redshifts
may depend on position.

\begin{figure}[htb!]
\centering
\includegraphics[width=\linewidth]{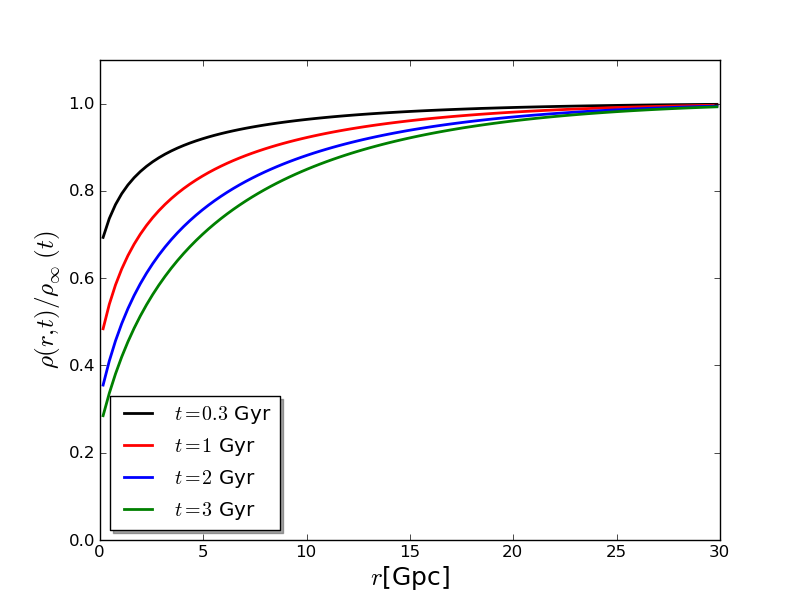} 
 \caption{The matter density relative to the matter density outside of the void at different potential
 galaxy formation times. The best-fit void parameters to the full set of data (SN + $H_0$ + Ages + $\Omega_m$) is used.
 Even at cosmic times as early as $t = 300$ Myr, the void underdensity is significant, with a $\sim 30 \%$ underdensity at the center.
  }
  \label{fig:tfprofile}
\end{figure}

How significant this position dependence is, is determined by the void profile at the time of galaxy formation.
If at these early times the void is reduced to a small perturbation onto a homogeneous universe (which in the $t \to 0$ limit is the case), any spatial
variation of formation time can likely be neglected, but if the void still represents a significant underdensity,
this variation may render the galaxy age-based Hubble measurement invalid.
Since we do not know the typical formation times of the galaxies defining the red envelope,
we consider in Fig.~\ref{fig:tfprofile} a range of cosmic times\footnote{To put these times in context, the average best-fit value of $t_{f,i}$ based on the
full SN + $H_0$ + Ages + $\Omega_m$ data in the void model is about $1$ Gyr, while in the $\Lambda CDM$ case,
we find $2.4$ Gyr. A previous study of the effect of inhomogeneous formation time, \cite{wangzhang12}, considered $0.3$ Gyr as a plausible formation time.}.
The figure shows the void profile, quantified by the matter density relative to the matter density outside of the void, for the best-fit void profile to
the data (SN + $H_0$ + Ages + $\Omega_m$), and shows that, even if typical formation times are as low as $0.3$ Gyr, the underdensity
is still approximately $30 \%$ in the void center. The later the formation time, the deeper the void is.

We next study the effect on the Hubble parameter determination more quantitatively.
Using
\beq
t = t_f + {\rm Age},
\eeq
where $t$ is cosmic time and $t_f$ the formation time, the usual estimator for the Hubble parameter would lead to a bias
if $t_f = t_f(r)$ because
\beq
- \frac{d{\rm Age}}{d\ln(1 + z)} = H_L^{-1} + \frac{dt_f(r(z))}{d\ln(1 + z)}.
\eeq
Thus, if $t_f(r)$ is known, one can estimate the relative bias in the (inverse) Hubble parameter,
\beq
\label{eq:bias}
\frac{\Delta H^{-1}_L}{H^{-1}_L} = H_L \, \frac{dt_f(r(z))}{d\ln(1 + z)}.
\eeq

\begin{figure}[htb!]
\centering
\includegraphics[width=\linewidth]{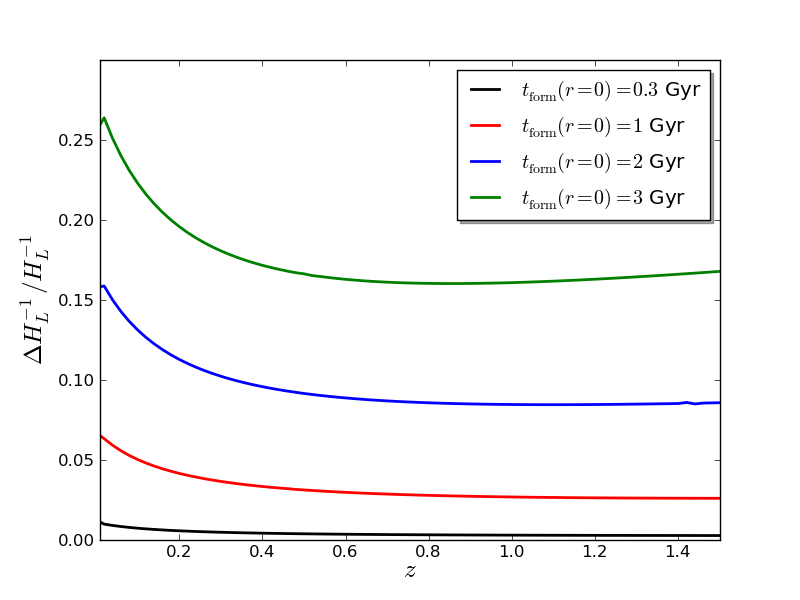} 
 \caption{The error in the determination of the (inverse of the) longitudinal Hubble parameter
 due to assuming the statistics of the galaxy formation time to be independent of location in the void.
 The calculation uses the best-fit void model to the SN + $H_0$ + Ages + $\Omega_m$ data set, and a simple
 model that assumes the galaxy formation time $t_f$ is reached when some critical energy density is reached (see text for details).
 Depending on the time of formation in the center of the void, the error in $H_L$ can be large, up to $15 \%$ at low redshift if the formation
 time is $t_f = 1$ Gyr.
  }
  \label{fig:tfHz}
\end{figure}

Since we do not have a good way of calculating $t_f(r)$, we consider a toy model, which, while not quantitatively
correct, will at least tell us whether or not the bias is likely to be significant.
The toy model simply assumes that $t_f(r)$ is reached when the local matter density reaches a critical value $\rho_*$,
i.e.~
\beq
\rho(r, t_f(r)) = \rho_*.
\eeq
This results in
\beq
\frac{dt_f}{d\ln(1 + z)} = - \frac{dr}{d\ln(1 + z)} \, \left[\frac{\pa \rho/\pa r}{\pa \rho/\pa t}\right]_{t = t_f(r)},
\eeq
which, together with Eq.~(\ref{eq:bias}), allows the calculation of $\Delta H^{-1}_L/H^{-1}_L$ once a void model is specified.
We calculate this bias, assuming the best-fit void model (again, to SN + $H_0$ + Ages + $\Omega_m$), for several choices of the formation time
at the center of the void, and show the results in Fig.~\ref{fig:tfHz}.
The bias in the toy model clearly depends strongly on the central formation time, with a formation time of $0.3$ Gyr leading to
a negligible bias compared to current error bars, but a formation time of $3$ Gyr causing a bias of up to $\sim 25 \%$.

Taking into account the large uncertainty in the central formation time, and in the theoretical evaluation of the spatial variation $t_f(r)$ (our estimate being based on a mere toy model),
we conclude that the bias in the Hubble parameter may well be significant and it is not safe to use the method by \cite{jimloeb02} without taking the effect
of inhomogeneous formation time into account. For this reason, we have in this work only used the red envelope ages as lower bounds on the age of the universe,
without estimating $H_L$. If a better understanding of the formation of galaxies in void cosmologies
were to lead to accurate predictions of $t_f(r)$ (perhaps with a small number of free parameters),
the information on $H_L$ could be partially or fully restored. In fact, in our toy model,
the effect of $t_f(r)$ on the differential age measurement is (void) cosmology dependent so that
the spatial variation of formation time would potentially even add new information.

\clearpage
\bibliographystyle{ieeetr}
\bibliography{refs}
\end{document}